\documentstyle[11pt,epsf]{article}

\newcommand{\ct}{\cite}
\newcommand{\lb}{\label}

\newcommand{\bc}{\begin{center}}
\newcommand{\ec}{\end{center}}
\newcommand{\bd}{\begin{displaymath}}
\newcommand{\ed}{\end{displaymath}}
\newcommand{\be}{\begin{equation}}
\newcommand{\ee}{\end{equation}}
\newcommand{\ba}{\begin{array}}
\newcommand{\ea}{\end{array}}
\newcommand{\bea}{\begin{eqnarray}}
\newcommand{\eea}{\end{eqnarray}}
\newcommand{\bt}{\begin{tabular}}
\newcommand{\et}{\end{tabular}}
\newcommand{\un}{\underline}
\newcommand{\ov}{\overline}
\newcommand{\bp}{\begin{picture}}
\newcommand{\ep}{\end{picture}}
\newcommand{\bfi}{\begin{figure}}
\newcommand{\efi}{\end{figure}}
\newcommand{\ds}{\displaystyle}

\def\fun#1#2{\lower3.6pt\vbox{\baselineskip0pt\lineskip.9pt
\ialign{$\mathsurround=0pt#1\hfil##\hfil$\crcr#2\crcr\sim\crcr}}}

\begin{document}

\vspace{3cm}

\title{\Large\bf {Phase Transition Couplings in U(1) and SU(N) Regulirized
Gauge Theories}}
\author{{\bf L.V.Laperashvili}
\footnote{{\bf E-mail}:laper@heron.itep.ru}\\
\it Institute of Theoretical and Experimental Physics,\\
\it B.Cheremushkinskaya 25, 117218 Moscow, Russia \\[0.2cm]
{\bf H.B.Nielsen}
\footnote{{\bf E-mail}:hbech@alf.nbi.dk}\\
\it Niels Bohr Institute,\\
\it DK-2100 Copenhagen {\O}, Denmark\\[0.2cm]
{\bf D.A.Ryzhikh}
\footnote{{\bf E-mail}:ryzhikh@heron.itep.ru}\\
\it Institute of Theoretical and Experimental Physics,\\
\it B.Cheremushkinskaya 25, 117218 Moscow, Russia}

\date{}

\maketitle
\thispagestyle{empty}

\vspace{1cm}

{\large
monopoles, loops\\

\vspace{3cm}

\un{Corresponding author:} \\
Prof.H.B.Nielsen,\\
Niels Bohr Institute \\
Blegdamsvej 17\\
DK-2100 Copenhagen {\O}\\
Denmark\\
Telephone: + 45 353 25259\\
E-mail: hbech@alf.nbi.dk}\\

\vspace*{-21cm}
\begin{flushright}
{\Large Preprint NBI--HE--01--07~~~~~}
\end{flushright}

\newpage
\thispagestyle{empty}
\vspace{10cm}
\begin{abstract}

Using a 2--loop approximation for $\beta$--functions, we have considered
the corresponding renormalization group improved effective potential in
the Dual Abelian Higgs Model (DAHM) of scalar monopoles and calculated
the phase transition (critical) couplings in U(1) and SU(N) regularized
gauge theories. In contrast to our previous result $\alpha_{crit}
\approx 0.17$, obtained in the one--loop approximation with the DAHM
effective potential (see Ref.\ct{19}), the critical value of the electric
fine structure constant in the 2--loop approximation, calculated in the
present paper, is equal to $\alpha_{crit}\approx 0.208$ and coincides
with the lattice result for compact QED \ct{10}: $\alpha_{crit}^{lat}
\approx 0.20\pm 0.015$.
Following the 't Hooft's idea of the "abelization" of monopole vacuum in
the Yang--Mills theories, we have obtained an estimation of the SU(N)
triple point coupling constants, which is $\alpha_{N,crit}^{-1}
=\frac{N}{2}\sqrt{\frac{N+1}{N-1}}\,\alpha_{U(1),crit}^{-1}$. This relation
was used for the description of the Planck scale values of the inverse
running constants $\alpha_i^{-1}(\mu)$  (i=1,2,3 correspond to U(1), SU(2)
and SU(3) groups), according to the ideas of the Multiple Point Model \ct{15}.

\end{abstract}

\newpage

\pagenumbering{arabic}
\vspace{1cm}
\large
\section{Introduction}
\vspace{0.51cm}

Lattice gauge theories, first introduced by K.Wilson \ct{1} for studying
the problem of confinement, are described by the following simplest action:
\be
      S = - \frac{\beta}{N}\sum_p Re(Tr\,{\cal U}_p),          \lb{1}
\ee
where the sum runs over all plaquettes of a hypercubic lattice and
${\cal U}_p$ is the product around the plaquette $p$ of the link
variables in the N--dimensional fundamental representation of the
gauge group G. Monte Carlo simulations of these simple Wilson lattice
theories in 4 dimensions showed a (or an almost) second--order
deconfining phase transition for U(1) \ct{2},\ct{3}, a crossover
behavior for SU(2) and SU(3) \ct{4},\ct{5}, and a first--order
phase transition for SU(N) with $N\ge 4$ \ct{6}. Bhanot and
Creutz \ct{7},\ct{8} have generalized the simple Wilson action,
introducing two parameters in action:
\be
   S = \sum_p[-\frac{\beta_f}{N}Re(Tr\,{\cal U}_p) -
               \frac{\beta_A}{N^2-1}Re(Tr_A{\cal U}_p)],   \lb{2}
\ee
where $Tr_A$ is the trace in the adjoint representation of SU(N).
The phase diagrams, obtained for the generalized lattice SU(2) and
SU(3) theories (\ref{2}) by Monte Carlo methods in Refs.\ct{7},\ct{8},
showed the existence of a triple point which is a boundary point of three
first--order phase transitions: the "Coulomb--like" and $SU(N)/Z_N$
and $Z_N$ confinement phases meet together at this point. From the triple
point emanate three phase border lines which separate the corresponding
phases. The $Z_N$ phase transition is a "discreteness" transition, occurring
when lattice plaquettes jump from the identity to nearby elements
in the group. The $SU(N)/Z_N$ phase transition is due to a condensation
of monopoles (a consequence of the non-trivial $\Pi_1$ of the group).

Monte Carlo simulations of the U(1) gauge theory, described by the
two-parameter lattice action \ct{9},\ct{10}:
\be
     S = \sum_p[\beta^{lat} \cos \Theta_p + \gamma^{lat} \cos2\Theta_p],\quad
     {\mbox {where}} \quad {\cal U}_p = e^{i\Theta_p},     \lb{3}
\ee
also indicate the existence of a triple point on the corresponding
phase diagram: "Coulomb--like", totally confining and $Z_2$ confining
phases come together at this triple point.

The next efforts of the lattice simulations of the SU(N) gauge theories
may be found in the review \ct{10a}.

The lattice artifact monopoles are responsible for the confinement
mechanism in lattice gauge theories what is confirmed by many
numerical and theoretical investigations (see reviews \ct{11} and papers
\ct{12}). The simplest effective dynamics describing the
confinement mechanism in the pure gauge lattice U(1) theory
is the Dual Abelian Higgs Model (DAHM) of scalar monopoles \ct{13}.

As it was shown in a number of investigations (see \ct{11},\ct{12}
and references there), the confinement in the SU(N) lattice gauge theories
effectively comes to the same U(1) formalism. The reason is the Abelian
dominance in their monopole vacuum: monopoles of the Yang--Mills theory are
the solutions of the U(1)--subgroups, arbitrary embedded into the SU(N) group.
After a partial gauge fixing (Abelian projection by 't Hooft \ct{14}) SU(N)
gauge theory is reduced to an Abelian $U(1)^{N-1}$ theory with $N-1$
different types of Abelian monopoles.  Choosing the Abelian gauge
for dual gluons, it is possible to describe the confinement in the lattice
SU(N) gauge theories by the analogous dual Abelian Higgs model of scalar
monopoles.

In our previous papers \ct{15}-\ct{17} the calculations of the U(1)
phase transition (critical) coupling constant were connected with the
existence of artifact monopoles in the lattice gauge theory and also
in the Wilson loop action model \ct{17}.
But in Refs.\ct{18},\ct{19} we have considered the Higgs monopole model
approximating the lattice artifact monopoles as fundamental pointlike
particles described by Higgs scalar field. Using a one--loop renormalization
group improvement of the effective Coleman--Weinberg potential \ct{20} written
for the dual sector of scalar electrodynamics, we have calculated in \ct{19}
the U(1) critical values of the magnetic fine structure constant ${\tilde
\alpha}_{crit} = g^2_{crit}/4\pi\approx 1.48$ and electric fine structure
constant $\alpha_{crit} = \pi/g^2_{crit} \approx {0.17}$ (by the Dirac
relation). These values are very close to the lattice result \ct{10} for the
compact QED described by the simple Wilson action corresponding to the
case $\gamma^{lat}=0$ in Eq.(\ref{3}):
\be
\alpha_{crit}^{lat}\approx 0.20\pm 0.015\quad
{\mbox{and}} \quad {\tilde \alpha}_{crit}^{lat}\approx 1.25\pm 0.10.
                   \lb{4}
\ee

In the present paper we have considered a two--loop approximation
for the renormalization group improved effective potential in the same
DAHM of scalar monopoles (Section 2).
We have obtained the following
result: $\alpha_{crit}\approx 0.208$, which coincides with the lattice
result (\ref{4}).

Using the idea of the "abelization" of the monopole vacuum in the
SU(N) lattice gauge theories, we have developed a method of
theoretical estimation of the SU(N) critical couplings (see Section 3).

Investigating the phase transition in the dual Higgs monopole model,
we have pursued two objects. From one side, we had an aim to
explain the lattice results. But we had also another aim.

According to the Multiple Point Model (MPM) \ct{15}, there is a multiple
critical point at the Planck scale where vacua of all fields existing
in Nature are degenerate. Having at the Planck scale the phase transitions
in U(1), SU(2) and SU(3) sectors of the fundamental regularized
gauge theory, it is natural to assume that the objects responsible for
these transitions are the physically existing Higgs scalar monopoles
which have to be introduced into theory as fundamental fields.
Our present calculations indicate that the corresponding critical
couplings coincide with the lattice ones, confirming the idea of
Ref.\ct{15}. Sections 4 and 6 are devoted to this problem.

\section{Higgs monopole model renormalization group equations\\
with two-loop approximation for beta-functions}

As it was mentioned in Introduction, the DAHM of scalar
monopoles describes the dynamics of confinement in lattice theories.
This model, first suggested in Refs.\ct{13}, considers the
following Lagrangian:
\be
    L = - \frac{1}{4g^2} F_{\mu\nu}^2(B) + \frac{1}{2} |(\partial_{\mu} -
           iB_{\mu})\Phi|^2 - U(\Phi),              \lb{5}
\ee
where
\be
 U(\Phi) = \frac{1}{2}\mu^2 {|\Phi|}^2 + \frac{\lambda}{4}{|\Phi|}^4
                                            \lb{6}
\ee
is the Higgs potential of scalar monopoles with magnetic charge $g$, and
$B_{\mu}$ is the dual gauge field interacting with the monopole field $\Phi$.
In this model $\mu^2$ is negative.

In Eqs.(\ref{5}),(\ref{6}) the complex scalar field $\Phi$ contains
the Higgs ($\phi$) and Goldstone ($\chi$) boson fields:
\be
          \Phi = \phi + i\chi.             \lb{7}
\ee

The effective potential in the Higgs model of scalar electrodynamics
was first calculated by Coleman and Weinberg \ct{20} in the one--loop
approximation. The general methods of its calculation is given in
review \ct{21}. The effective potential $V_{eff}(\phi_c)$ is a function of
the classical field $\phi_c$ \ct{20}:
\be
       V_{eff}(\phi_c) = \sum_n \frac{-1}{n!}\Gamma^{(n)}(0){\phi_c}^n,
                                           \lb{8}
\ee
where $\Gamma^{(n)}(0)$ is the one--particle irreducible n--point Green's
function calculated at zero external momenta.

The RGE for the effective potential means that the potential cannot
depend on a change in the arbitrary parameter -- renormalization scale $M$,
i.e. $dV_{eff}/dM = 0$. The effects of changing it are absorbed into
changes in the coupling constants, masses and fields, giving so--called
running quantities.

Considering the RG improvement of the effective potential \ct{20}, \ct{21}
and choosing the evolution variable as
\be
                   t = \log(\phi^2/M^2),       \lb{9}
\ee
we have the following RGE for the improved $V_{eff}(\phi^2)$
with $\phi^2\equiv \phi^2_c$ \ct{21a}:
\be
   (M^2\frac{\partial}{\partial M^2} + \beta_{\lambda}\frac{\partial}
{\partial\lambda} + \beta_g\frac{\partial}{\partial g^2} +
\beta_{(\mu^2)}\mu^2\frac{\partial}{\partial \mu^2} - \gamma\phi^2
\frac{\partial}{\partial \phi^2})V_{eff}(\phi^2) = 0,    \lb{10}
\ee
where $\gamma$ is the anomalous dimension and $\beta_{(\mu^2)}$,
$\beta_{\lambda}$ and $\beta_g$ are the RG $\beta$--functions for mass,
scalar and gauge couplings, respectively. RGE (\ref{10}) leads to the
following form of the improved effective potential \ct{20}:
\be
     V_{eff} = \frac{1}{2}\mu^2_{run}(t)G^2(t)\phi^2 +
                 \frac{1}{4}\lambda_{run}(t)G^4(t)\phi^4,   \lb{11}
\ee
where $\mu^2_{run}$ and $\lambda_{run}$ are the running squared mass
of scalar monopoles and their self--interaction constant, respectively,
and (in our notations):
\be
 G(t) = \exp[-\frac{1}{2}\int_0^t dt'\,\gamma\left(g_{run}(t'),
         \lambda_{run}(t')\right)].                              \lb{12}
\ee
A set of ordinary differential equations (RGE) corresponds to Eq.(\ref{10}):

\be
    \frac{d\lambda_{run}}{dt} = \beta_{\lambda}\left(g_{run}(t),
                               \lambda_{run}(t)\right),           \lb{13}
\ee
\be
    \frac{d\mu^2_{run}}{dt} = \mu^2_{run}(t)\beta_{(\mu^2)}
             \left(g_{run}(t), \lambda_{run}(t)\right),           \lb{14}
\ee
\be
    \frac{dg^2_{run}}{dt} = \beta_g\left(g_{run}(t),
                               \lambda_{run}(t)\right).           \lb{15}
\ee
So far as the mathematical structure of the DAHM is equivalent
to the Higgs scalar electrodynamics, we can use all results of the last
theory in our calculations, replacing the electric charge $e$ and photon
field $A_{\mu}$ by magnetic charge $g$ and dual gauge field $B_{\mu}$.

The RG $\beta$--functions for different renormalizable gauge theories with
semisimple group have been calculated in the two--loop approximation
\ct{22}-\ct{27} and even beyond \ct{28}. But in this paper we made use
the results of Refs.\ct{22} and \ct{25} for calculation of $\beta$--functions
and anomalous dimension in the two--loop approximation, applied to the
DAHM with scalar fields (\ref{7}). The higher approximations essentially
depend on the renormalization scheme \ct{28}.
Thus, on the level of two--loop approximation we have:
\be
             \beta_{\lambda} =
                       \beta_{\lambda}^{(1)} +
                           \beta_{\lambda}^{(2)},              \lb{16}
\ee
where
\be
\beta_{\lambda}^{(1)} = \frac{1}{16\pi^2}(10\lambda^2 - 6g^2\lambda + 3g^4)
                                                        \lb{17}
\ee
and
\be
  \beta_{\lambda}^{(2)} = \frac{1}{(16\pi^2)^2}( - 25\lambda^3 +
   \frac{15}{2}g^2{\lambda}^2 - \frac{229}{12}g^4\lambda - \frac{59}{6}g^6).
                                                               \lb{18}
\ee
The corresponding $\beta$--function for the squared running mass is:
\be
   \beta_{(\mu^2)} = \beta_{(\mu^2)}^{(1)} + \beta_{(\mu^2)}^{(2)},
                                                               \lb{19}
\ee
where
\be
\beta_{(\mu^2)}^{(1)} = - \frac{3g^2}{16\pi^2} + \frac{\lambda}{4\pi^2}
                                                               \lb{20}
\ee
and
\be
\beta_{(\mu^2)}^{(2)} = \frac{1}{(16\pi^2)^2}(\frac{31}{12}g^4 + 3\lambda^2).
                                                                 \lb{21}
\ee
The gauge coupling $\beta$--function is given by Ref.\ct{22}:
\be
     \beta_g = \beta_g^{(1)} + \beta_g^{(2)} =
             \frac{g^4}{48\pi^2} + \frac{g^6}{(16\pi^2)^2}.
                                                                 \lb{22}
\ee
Anomalous dimension follows from calculations made in Ref.\ct{25}:
\be
    \gamma = \gamma^{(1)} + \gamma^{(2)} =
           - \frac{3g^2}{16\pi^2} + \frac{1}{(16\pi^2)^2}\frac{31}{12}g^4.
                                                                 \lb{23}
\ee
In Eqs.(\ref{16}-\ref{23}) and below, for simplicity, we have used the
following notations: $\lambda\equiv \lambda_{run}$, $g\equiv g_{run}$ and
$\mu\equiv \mu_{run}$.

\section{Calculation of the U(1) critical couplings in the Higgs monopole
model using renormalization group equations}

The effective potential (\ref{11}) can have several minima. Their positions
depend on $g^2$, $\mu^2$ and $\lambda$. If the first local minimum occurs
at $\phi = 0$ and $V_{eff}(0) = 0$, it corresponds to the Coulomb--like phase.
In the case when the effective potential has the second local minimum at
$\phi = \phi_{min} \neq 0\,$ with $\,V_{eff}^{min}(\phi_{min}^2) < 0$,
we have the confinement phase. The phase transition between the
"Coulomb--like" and confinement phases is given by the condition when
the first local minimum
at $\phi = 0$ is degenerate with the second minimum at $\phi = \phi_0$.
These degenerate minima are shown in Fig.1 by the curve 1. They correspond
to the different vacua arising in this model. And the dashed curve 2
describes the appearance of two minima corresponding to the confinement
phases (see details in Ref.\ct{19}).

The conditions of the existence of degenerate vacua are given by the
following equations:
\be
           V_{eff}(0) = V_{eff}(\phi_0^2) = 0,     \lb{24}
\ee
\be
    \frac{\partial V_{eff}}{\partial \phi}|_{\phi=0} =
    \frac{\partial V_{eff}}{\partial \phi}|_{\phi=\phi_0} = 0,   \lb{25}
\ee
and inequalities
\be
    \frac{\partial^2 V_{eff}}{\partial \phi^2}|_{\phi=0} > 0, \qquad
    \frac{\partial^2 V_{eff}}{\partial \phi^2}|_{\phi=\phi_0} > 0.
                                                                  \lb{26}
\ee
The first equation (\ref{24}) applied to Eq.(\ref{11}) gives:
\be
    \mu^2_{run} = - \frac{1}{2} \lambda_{run}(t_0)\,\phi_0^2\, G^2(t_0), \lb{27}
\ee
where $t_0 = \log(\phi_0^2/M^2)$.

Using a notation:
\be
            V'_{eff}(\phi_0^2)\equiv
    \frac{\partial V_{eff}}{\partial \phi^2}|_{\phi=\phi_0},     \lb{28}
\ee
we have the condition:
\be
            V'_{eff}(\phi_0^2) = 0,                     \lb{29}
\ee
according to the second equation in (\ref{25}).
Calculating the first derivative of $V_{eff}$, we obtain the following
expression:
$$
 V'_{eff}(\phi^2) = \frac{V_{eff}(\phi^2)}{\phi^2}(1 + 2\frac{d\log G}{dt}) +
               \frac 12 \frac{d\mu^2_{run}}{dt} G^2(t)
$$
\be
  + \frac 14 \biggl(\lambda_{run}(t) + \frac{d\lambda_{run}}{dt} +
      2\lambda_{run}\frac{d\log G}{dt}\biggr)G^4(t)\phi^2.
                     \lb{30}
\ee
From Eq.(\ref{12}), we have:
\be
          \frac{d\log G}{dt} = - \frac{1}{2}\gamma .             \lb{31}
\ee
It is easy to find the joint solution of equations
\be
      V_{eff}(\phi_0^2) = V'_{eff}(\phi_0^2) = 0.         \lb{32}
\ee
Using RGE (\ref{13}), (\ref{14}) and Eqs.(\ref{27}), (\ref{31}), we obtain:
\be
 V'_{eff}(\phi_0^2) =\frac{1}{4}( - \lambda_{run}\beta_{(\mu^2)} +
\lambda_{run} + \beta_{\lambda} - \gamma \lambda_{run})G^4(t_0)\phi_0^2 = 0,
                                                            \lb{33}
\ee
or
\be
    \beta_{\lambda} + \lambda_{run}(1 - \gamma - \beta_{(\mu^2)}) = 0.
                                                \lb{34}
\ee
Substituting $\beta_{\lambda},\,\beta_{(\mu^2)}$ and $\gamma$,
given by Eqs.(\ref{16}--\ref{21}) and (\ref{23}), we obtain from
Eq.(\ref{34}) the following equation:
\be
 3y^2 - 16\pi^2 + 6x^2 + \frac{1}{16\pi^2}(28x^3 + \frac{15}{2}x^2y +
  \frac{97}{4}xy^2 - \frac{59}{6}y^3) = 0,             \lb{35}
\ee
where $x = - \lambda_{PT}$ and $y = g^2_{PT}$ are the phase transition
values of $ - \lambda_{run}$ and $g^2_{run}$.
The curve (\ref{35}) describes a border between the "Coulomb--like"
($V_{eff} \ge 0$) and confinement ($V_{eff}^{min} < 0$) phases.
Choosing a physical branch corresponding to $g^2 \ge 0$ and $g^2\to 0$,
when $\lambda \to 0$, we have received curve 2 on the phase diagram
$(\lambda_{run}; g^2_{run})$ shown in Fig.2. This curve
corresponds to the 2--loop approximation and can be compared with
the  curve 1 of Fig.2, which describes the same phase border
calculated in Ref.\ct{19} in the 1--loop approximation. In the previous
paper \ct{19} we have emphasized that our accuracy of the 1--loop
approximation is not excellent and can commit errors of order 30\%.
Curves 1,2 resemble this situation.

According to the phase diagram drawn in Fig.2, the confinement phase
begins at $g^2 = g^2_{max}$ (see details in Ref.\ct{19}) and exists at
$g^2 \le g^2_{max}$. Therefore, we have:
$$
  g^2_{crit} = g^2_{max1}\approx 18.61
\quad -\quad {\mbox{in the 1-loop approximation}},
$$
and
\be
  g^2_{crit} = g^2_{max2}\approx 15.11
\quad -\quad {\mbox{in the 2-loop approximation}}.
                                                      \lb{36}
\ee
We see the deviation of results of order 20\%.

In the 2--loop approximation:
\be
   \tilde \alpha_{crit} = \frac {g^2_{crit}}{4\pi}\approx 1.20,
                                                             \lb{37}
\ee
instead of the result $\tilde \alpha_{crit} \approx 1.48$ obtained in the
1--loop approximation.

Using the Dirac relation for elementary charges:
\be
   eg = 2\pi, \quad{\mbox{or}}\quad \alpha \tilde \alpha = \frac{1}{4},
                                                       \lb{38}
\ee
we get a 2-loop approximation result for the critical electric fine
structure constant:
\be
        \alpha_{crit} = \frac{1}{4{\tilde \alpha}_{crit}}\approx 0.208.
                                                                \lb{39}
\ee
This result coincides with the lattice result (\ref{4}) obtained for the
compact QED by Monte Carlo methods \ct{10}. Eqs.(\ref{4}) and (\ref{39})
give the following result for the inverse electric fine structure
constant:
\be
          \alpha_{crit}^{-1}\approx 5.                   \lb{39a}
\ee
This value is important for the phase transition at the Planck scale
(see Sections 4 and 6).
Now we are able to estimate the validity of 2--loop approximation
for all $\beta$--functions and $\gamma$, calculating the corresponding
ratios of 2--loop contributions to 1--loop contributions
at the maxima of curves 1 and 2:
\be
\begin{array}{|l|l|}
\hline %
&\\[-0.2cm]%
\lambda_{crit} = \lambda_{run}^{max1} \approx{-13.16}&\lambda_{crit} =
\lambda_{run}^{max2}\approx{-7.13}\\[0.5cm]
g^2_{crit} = g^2_{max1}\approx{18.61}& g^2_{crit} = g^2_{max2}
\approx{15.11}\\[0.5cm]
\frac{\ds\gamma^{(2)}}{\ds\gamma^{(1)}}\approx{-0.0080}&\frac{\ds
\gamma^{(2)}}{\ds\gamma^{(1)}}\approx{-0.0065}\\[0.5cm]
\frac{\ds\beta_{\mu^2}^{(2)}}{\ds\beta_{\mu^2}^{(1)}}\approx{-0.0826}
&\frac{\ds\beta_{\mu^2}^{(2)}}{\ds\beta_{\mu^2}^{(1)}}
\approx{-0.0637}\\[0.8cm]
\frac{\ds\beta_{\lambda}^{(2)}}{\ds\beta_{\lambda}^{(1)}}\approx{0.1564}
&\frac{\ds\beta_{\lambda}^{(2)}}
{\ds\beta_{\lambda}^{(1)}}\approx{0.0412}\\[0.8cm]
\frac{\ds\beta_g^{(2)}}{\ds\beta_g^{(1)}}\approx{0.3536}&\frac{\ds
\beta_g^{(2)}}{\ds\beta_g^{(1)}}\approx{0.2871}\\[0.5cm]
\hline %
\end{array}
\lb{40}%
\ee %
Here we see that all ratios are sufficiently small, i.e. all
2--loop contributions are small in comparison with 1--loop contributions,
confirming the validity of perturbation theory in the 2--loop
approximation, considered in this model. The accuracy of deviation is worse
($\sim 30\%$) for $\beta_g$--function. But it is necessary to emphasize
that calculating the border curves 1 and 2 we have not used RGE (\ref{22})
for monopole charge: $\beta_g$--function is absent in Eq.(\ref{34}).
This means that we can expect a nice accuracy for $g^2_{crit}$
calculated with help of Eq.(\ref{35}).

The above--mentioned $\beta_g$--function appears only in the second order
derivative of $V_{eff}$ which is related with monopole mass $m$:
\be
               V''_{eff}(\phi_0^2) = \frac{1}{2\phi_0^2}\frac
{\partial^2V_{eff}}{\partial \phi^2}|_{\phi=\phi_0} =
                      \frac{m^2}{2\phi_0^2}.                  \lb{41}
\ee
As it was shown in Ref.\ct{19}, $V''_{eff}\to 0$ and monopole acquires
a zero mass near the critical point $(\lambda_{crit};\, g^2_{crit})$.
This result is in agreement with the result of compact QED described
by the Villain action \ct{29}: $m^2\approx 0$ in the vicinity of the
critical point.

\section{Multiple Point Model and critical values of the U(1)
and SU(N) fine structure constants}

A lot of investigations were devoted to the question: "What comes beyond
the Standard Model?". Grand Unification Theories (GUTs), unifying all
gauge interactions, were constructed and the role of supersymmetry in GUTs
was investigated as very promising. Unfortunately, at present time
experiment does not indicate any manifestation of the supersymmetry.
In this connection, the Anti--Grand Unification Theory (AGUT) was developed
in Refs.\ct{30}-\ct{35} as a realistic alternative to SUSY GUTs.
According to this theory, supersymmetry does not come into the existence
up to the Planck energy scale:
\be
             \mu_{Pl} = 1.22\cdot 10^{19}\,GeV.      \lb{42}
\ee
The Standard Model (SM) is based on the group:
\be
   SMG = SU(3)_c \otimes SU(2)_L \otimes U(1)_Y.      \lb{43}
\ee
The AGUT suggests that at the scale $\mu_G\sim \mu_{Pl}$ there exists
the more fundamental group $G$ containing $N_{gen}$ copies of the
Standard Model Group SMG:
\be
 G = SMG_1\otimes SMG_2\otimes...\otimes SMG_{N_{gen}}\equiv (SMG)^{N_{gen}},
                                               \lb{44}
\ee
where $N_{gen}$ designates the number of quark and lepton generations.

If $N_{gen}=3$ (as AGUT predicts), then the fundamental gauge group G is:
\be
    G = (SMG)^3 = SMG_1\otimes SMG_2\otimes SMG_3,       \lb{45}
\ee
or the generalized one:
\be
         G_f = (SMG)^3\otimes U(1)_f,             \lb{46}
\ee
which was suggested by the fitting of fermion masses of the SM
(see Refs.\ct{32}).

Recently a new generalization of the AGUT was suggested in Refs.\ct{34}:
\be
           G_{\mbox{ext}} = (SMG\otimes U(1)_{B-L})^3,    \lb{47}
\ee
which takes into account the see--saw mechanism with right-handed neutrinos,
also gives the reasonable fitting of the SM fermion masses and describes
all neutrino experiments known today.

The AGUT approach is used in conjunction with the Multiple Point Principle
(MPP) proposed in Ref.\ct{15}. According to MPP, there is a special point
--- the Multiple Critical Point (MCP) --- on the phase diagram of the
fundamental regularized gauge theory $G$ (or $G_f$, or $G_{\mbox{ext}}$),
which is a point where the vacua of all fields existing in Nature are
degenerate, having the same vacuum energy density. Such a phase diagram
has axes given by all coupling constants considered in theory.
Then all (or just many) numbers of phases meet at the MCP.

Multiple Point Model assumes the existence of MCP at the Planck scale,
insofar as gravity may be "critical" at the Planck scale (in the sense of
degenerate vacua).

The usual definition of the SM coupling constants:
\be
  \alpha_1 = \frac{5}{3}\frac{\alpha}{\cos^2\Theta_{\ov{MS}}},\quad
  \alpha_2 = \frac{\alpha}{\sin^2\Theta_{\ov{MS}}},\quad
  \alpha_3 \equiv \alpha_s = \frac {g^2_s}{4\pi},        \lb{48}
\ee
where $\alpha$ and $\alpha_s$ are the electromagnetic and SU(3)
fine structure constants, respectively, is given in the Modified
minimal subtraction scheme ($\ov{MS}$). Using RGE with experimentally
established parameters, it is possible to extrapolate the experimental
values of three inverse running constants $\alpha_i^{-1}(\mu)$
(here $\mu$ is an energy scale and i=1,2,3 correspond to U(1),
SU(2) and SU(3) groups of the SM) from Electroweak scale to the Planck
scale. The precision of the LEP data allows to make this extrapolation
with small errors (see \ct{36}). Assuming that these RGEs for
$\alpha_i^{-1}(\mu)$ are contingent not encountering new particles
up to $\mu\approx \mu_{Pl}$ and doing the extrapolation with one
Higgs doublet under the assumption of a "desert", the following results
for the inverses $\alpha_{Y,2,3}^{-1}$ (here $\alpha_Y\equiv \frac{3}{5}
\alpha_1$) were obtained in Ref.\ct{15} (compare with \ct{36}):
\be
   \alpha_Y^{-1}(\mu_{Pl})\approx 55.5; \quad
   \alpha_2^{-1}(\mu_{Pl})\approx 49.5; \quad
   \alpha_3^{-1}(\mu_{Pl})\approx 54.0.
                                                            \lb{49}
\ee
The extrapolation of $\alpha_{Y,2,3}^{-1}(\mu)$ up to the point
$\mu=\mu_{Pl}$ is shown in Fig.3.

According to the AGUT, at some point $\mu=\mu_G < \mu_{Pl}$ (but near
$\mu_{Pl}$) the fundamental group $G$ (or $G_f$, or $G_{\mbox{ext}}$)
undergoes spontaneous breakdown to the diagonal subgroup:
\be
      G \longrightarrow G_{diag.subgr.} = \{g,g,g || g\in SMG\},
                                                              \lb{50}
\ee
which is identified with the usual (lowenergy) group SMG.
The point $\mu_G\sim 10^{18}$ GeV also is shown in Fig.3, together with
a region of G--theory where AGUT works.

The AGUT prediction of the values of $\alpha_i(\mu)$ at $\mu=\mu_{Pl}$
is based on the MPP, which gives these values in terms of the
critical couplings $\alpha_{i,crit}$ \ct{30}-\ct{35}:
\be
            \alpha_i(\mu_{Pl}) = \frac {\alpha_{i,crit}}{N_{gen}}
                       = \frac{\alpha_{i,crit}}{3}          \lb{51}
\ee
for i=2,3 and
\be
\alpha_1(\mu_{Pl}) = \frac{\alpha_{1,crit}}{\frac{1}{2}N_{gen}(N_{gen} + 1)}
                   = \frac{\alpha_{1,crit}}{6}                \lb{52}
\ee
for U(1).

In Eqs.(\ref{51}) and (\ref{52}) $\alpha_{i,crit}$ are the triple point
values of the effective fine structure constants given by the
generalized lattice SU(3)--, SU(2)-- and U(1)-- gauge theories
described by Eqs.(\ref{2}) and (\ref{3}).

The authors of Refs.\ct{7}-\ct{10} were not able to obtain the lattice
triple point values of $\alpha_{i,crit}$ by the Monte Carlo simulation
methods. These values were calculated theoretically in Ref.\ct{15}.
Using the lattice \ct{7}-\ct{10} triple point values of
$(\beta_A;\,\beta_f)$ and $(\beta^{lat};\,\gamma^{lat})$, the authors of
Ref.\ct{15} have obtained $\alpha_{i,crit}$ by so--called "Parisi improvement
method" (see details in \ct{15}):
\be
    \alpha_{Y,crit}^{-1}\approx 9.2\pm 1,
    \quad \alpha_{2,crit}^{-1}\approx 16.5\pm 1, \quad
    \alpha_{3,crit}^{-1}\approx 18.9\pm 1.                 \lb{53}
\ee
According to MPM, assuming the existence of MCP at $\mu=\mu_{Pl}$, we have
the following prediction of AGUT, substituting the results (\ref{53})
in Eqs.(\ref{51}) and (\ref{52}):
\be
   \alpha_Y^{-1}(\mu_{Pl})\approx 55\pm 6; \quad
   \alpha_2^{-1}(\mu_{Pl})\approx 49.5\pm 3; \quad
   \alpha_3^{-1}(\mu_{Pl})\approx 57.0\pm 3.
                                                            \lb{54}
\ee
These results coincide with the results (\ref{49}) obtained by
extrapolation of the experimental data to the Planck scale
in the framework of the pure SM (without any new particles).

But now we see, that the first value of Eq.(\ref{53}) gives
$\alpha_{Y,crit}^{-1}\approx 9$ at the triple
point of the generalized U(1) lattice theory, what is larger
than the simple lattice result theory and
Higgs monopole model result (\ref{39}), obtained in this paper.

The next step of our paper is to explain the relation between the values of
U(1) and SU(N) critical coupling constants (\ref{53}).


\section{Monopoles strength group dependence}

Lattice gauge theories have lattice artifact monopoles. We suppose that
only those lattice artifact monopoles are important for the phase
transition calculations which have the smallest monopole charges.
Let us consider the lattice gauge theory with the gauge group
$ SU(N)/Z_N $ as our main example.
That is to say, we consider the adjoint representation action and do not
distinguish link variables forming the same one multiplied by any element of
the center of the group. The group $ SU(N)/Z_N $ is not simply connected
and has the first homotopic group $\Pi_1(SU(N)/Z_N)$ equal to $Z_N$.
The lattice artifact monopole with the smallest magnetic charge may be
described as a three-cube ( or rather a chain of three-cubes, describing the
time track) from which radiates magnetic field corresponding to $U(1)$
subgroup of gauge group $SU(N)/Z_N$ with the shortest length insight of
this group but still homotopically non-trivial. In fact, this $U(1)$
subgroup is obtained by the exponentiating generator:
\be
\frac{``\lambda_8``}{2} = \frac {1}{\sqrt{2N(N-1)}}
\left(\begin{array}{*{4}{c}}
N-1 & 0 & \cdots & 0\\
0 & -1 & \cdots & 0 \\
\vdots & \vdots &\ddots & \vdots\\
0 & 0 &  \cdots & -1 \\
\end{array}
\right)                                      \lb{1a}
\ee
This specific form is one gauge choice; any similarity transformation
of this generator would describe physically the same monopole. If one
has somehow already chosen the gauge monopoles with different but similarity
transformation related generators, they would be physically different.
Thus, after gauge choice, there are monopoles corresponding to different
directions of the Lie algebra generators in the form ${\cal
U}\frac{``\lambda_8``}{2}\,{\cal U}^{+}$.

Now, when we  want to apply the effective potential calculation as a
technique for the getting phase diagram information for the condensation
of the lattice artifact monopoles in the non-abelian lattice gauge theory,
we have to correct the abelian case calculation for the fact that
after gauge choice we have a lot of different monopoles.
If a couple of monopoles happens to have their generators just in the same
directions in the Lie algebra, they will interact with each other as
Abelian monopoles (in first approximation). In general, the interaction
of two monopoles by exchange of a photon will be modified by the
following factor:

\be
  \frac{ Tr ({\cal U}_1\frac{``\lambda_8``}{2} {\cal U}_1^{+}\,
{\cal U}_2\frac{``\lambda_8``}{2} {\cal U}_2^{+})}
{Tr{(\frac{``\lambda_8``}2)}^2}.          \lb{2a}
\ee

We shall assume that we can correct these values of monopole orientations in
the Lie algebra in a statistical way. That is to say, we want to determine
an effective coupling constant ${\tilde g}_{eff}$ describing the
monopole charge as if there is only one Lie algebra orientationwise type
of monopole. It should be estimated statistically in terms of the monopolic
charge ${\tilde g}_{genuine}$ valid to describe the interaction between
monopoles with generators oriented along the same $U(1)$ subgroup.
A very crude intuitive estimate of the relation between these two
monopole charge concepts ${\tilde g}_{genuine}$ and ${\tilde g}_{eff}$
consists in playing that the generators are randomly oriented in the
whole $N^2 - 1$ dimensional Lie algebra. When even the sign of the Lie
algebra generator associated with the monopole is random --- as we assumed
in this crude argument --- the interaction between two monopoles with
just one photon exchanged averages out to zero. Therefore, we can get
a non-zero result only in the case of exchange by two photons or more.
That is, however, good enough for our effective potential calculation
since only ${\tilde g}^4$ (but not the second power) occurs in the
Coleman --- Weinberg effective potential in the one--loop approximation
(see \ct{19}-\ct{21}). Taking into account this fact that we can
average imagining monopoles with generators along a basis vector in the
Lie algebra, the chance of interaction by double photon exchange between
two different monopoles is just $\frac{1}{N^2 - 1}$, because there are
$N^2 - 1$ basis vectors in the basis of the Lie algebra. Thus, this crude
approximation gives:
\be
    {\tilde g}^4_{eff} = \frac{1}{N^2 - 1}{\tilde g}^4_{genuine}.
                                     \lb{3a}
\ee

Note that considering the two photons exchange which is forced by our
statistical description, we must concern the forth power
of the monopole charge $\tilde g$.

The relation (\ref{3a}) was not derived correctly, but its
validity can be confirmed if we use a more correct statistical argument.
The problem with our crude estimate is that the generators making monopole
charge to be minimal must go along the shortest type of U(1) subgroups with
non-trivial homotopy.

\subsection{Correct averaging}

The $\lambda_8$--like generators
${\cal U}\frac{``\lambda_8``}{2}\,{\cal U}^+$  maybe written as
\begin{equation}
{\cal U} \frac{``\lambda_8``}{2} \,{\cal U}^{+}
= - \sqrt{\frac {1}{2N(N-1)}} {\bf 1\!\!\! 1} + \sqrt{\frac{N}{2(N-1)}}
\,{\cal P}                                       \lb{4a}
\end{equation}
where $\cal P$ is a projection metrics into one--dimensional state in
the \un{N} representation. It is easy to see that averaging according to
the Haar measure distribution of $\cal U$, we get the average of
$\cal P$ projection on ``quark'' states with a distribution corresponding
to the rotationally invariant one on the unit sphere in the N--dimensional
\un{N}--Hilbert space.

If we denote the Hilbert vector describing the state
on which $\cal P$ shall project as
\be
{\cal P} =\left(
\begin{array}{c}
\psi_1\\
\psi_2\\
\vdots\\
\psi_N
\end{array}
\right),
                                       \lb{5a}
\ee
then the probability distribution on the unit sphere becomes:
\be
P (\left(
\begin{array}{c}
\psi_1\\
\psi_2\\
\vdots\\
\psi_N
\end{array}
\right)) {\prod}_{i=1}^N d{\psi}_i\propto \delta({\sum}_{i=1}^N {|{\psi}_i|}^2
- 1) \prod_{i=1}^N d({|\psi_i|}^2).               \lb{6a}
\ee
Since, of course, we must have ${|{\psi}_i|}^2 \ge 0$ for all $i=1,2,...,N$,
the $\delta$--function is easily seen to select a flat distribution
on a (N - 1)--dimensional equilateral simplex. The average of the two
photon exchange interaction given by the
correction factor (\ref{2a}) squared (numerically):
\begin{equation}
\frac{Tr ({\cal U}_1 \frac{``\lambda_8``}{2} {\cal U}_1^{+}{\cal U}_2
{\frac{``\lambda_8``}{2}}{\cal U}_2^{+})^2}{Tr({(\frac{``\lambda_8``}{2})}^2)
^2}
                                 \lb{7a}
\end{equation}
can obviously be replaced by the expression where we take as random only
one of the ``random``  $\lambda_8$--like generators, while the other one is
just taken as $\frac{``\lambda_8``}{2}$, i.e. we can take say
${\cal U}_2 = {\bf 1\!\!\!1}$ without changing the average.

Considering the two photon exchange diagram, we can write
the correction factor (obtained by the averaging)
for the fourth power of magnetic charge:
\be
\frac{{\tilde g}^4_{eff}}{{\tilde g}^4_{{\rm genuine}}}
= {\rm average}\{\frac{Tr{(\frac{``\lambda_8 ``}{2} {{\cal U}_{1}}
{\frac{``\lambda_8``}{2}}{{\cal U}_{1}}^{+})}^2}
{Tr{({(\frac{``\lambda_8``}{2})}^2)}^2}\}.       \lb{8a}
\ee
Substituting the expression (\ref{4a}) in Eq.(\ref{8a}), we have:
\be
\frac{{\tilde g}^4_{eff}}{{\tilde g}^4_{{\rm genuine}}}
= {\rm average}\{\frac{ Tr{(\frac{``\lambda_8``}{2} (-\sqrt{\frac{1}
{2N(N-1)}}{\bf 1\!\!\!1}
 + \sqrt{\frac {N}{2(N-1)}}{\cal P}))}^2}
{Tr{({(\frac{``\lambda_8``}{2})}^2)}^2}\}.
                                            \lb{8b}
\ee
Since $\frac{``\lambda_8``}{2}$ is traceless, we obtain using the
projection (\ref{5a}):

\be
 Tr(\frac{``\lambda_8``}{2} {\cal P}\sqrt{\frac{N}{2(N-1)}})
= - \frac{1}{2(N-1)} + \frac{N}{2(N-1)}{|\psi_1|}^2.
                                              \lb{9a}
\ee
The value of the square ${|\psi_1|}^2$ over the simplex is proportional
to one of the heights in this simplex. It is obvious from the geometry of a
simplex that the distribution of ${|\psi_1|}^2$ is
\be
     d{\mbox P} = (N-1){(1 - {|\psi_1|}^2)}^{(N-2)} d({|\psi_1|}^2),
                                          \lb{10a}
\ee
where, of course, $0 \le {|{\psi}_1|}^2 \le 1$ only is allowed.
In Eq.(\ref{10a}) P is a probability. By definition:
\be
     {\rm average}\{f({|{\psi}_1|}^2)\}
= (N-1)\int_0^1 f({|{\psi}_1|}^2)(1 - {|\psi_1|}^2)^{(N-2)}d({|\psi_1|}^2).
                                            \lb{11a}
\ee
Then
\begin{eqnarray}
\frac{{\tilde g}^4_{eff}}{{\tilde g}^4_{{\rm genuine}}}
&=& \frac{N^2}{(N-1)} \int_0^1
{(\frac{1}{N} - {|\psi_1|}^2)}^2{(1 - {|\psi_1|}^2)}^{N-2} d({|\psi_1|}^2)\\
&=& \frac{N^2}{N-1}\int_0^1 {(1-y-\frac{1}{N})}^2\,y^{N-2}dy\\
&=& \frac{1}{N^2-1}                \lb{12a}
\end{eqnarray}
and we have confirmed our crude estimation (\ref{3a}).

\subsection{Relative normalization of couplings}

Now we are interested in how ${\tilde g}^2_{genuine}$ is related to
$\alpha_N=g^2_N/{4\pi}$.

We would get the simple Dirac relation:
\be
          g_{(1)}\cdot {\tilde g}_{\rm{genuine}} = 2\pi
                                              \lb{13a}
\ee
if $g_{(1)}\equiv g_{U(1)-{\rm subgroup}}$ is the coupling for the
U(1)--subgroup of SU(N) normalized in such a way that the charge quantum
$g_{(1)}$ corresponds to a covariant derivative
$\partial_{\mu} - g_{(1)}\,A_{\mu}^{U(1)}$.

Now we shall follow the convention --- usually used to define
$\alpha_N={g^2_N}/{4\pi}$ --- that the covariant derivative for the
{\un{N}}--plet representation is:
\be
        D_{\mu} = \partial_{\mu} - g_N \frac{\lambda^a}{2}A_{\mu}^a
                                                     \lb{14a}
\ee
with
\be
 Tr(\frac{\lambda^a}{2}\frac{\lambda^b}{2}) = \frac{1}{2}\delta^{ab},
                                                   \lb{15a}
\ee
and the kinetic term for the gauge field is
\be
       L = - \frac{1}{4}F_{\mu\nu}^aF^{b\,\mu\nu},   \lb{16a}
\ee
where
\be
    F_{\mu\nu}^a = \partial_{\mu}A_{\nu}^a - \partial_{\nu}A_{\mu}^a
                    - g_N\,f^{abc}A_{\mu}^bA_{\nu}^c.
                                         \lb{17a}
\ee
Especially if we want to choose a basis for our generalized
Gell--Mann matrices so that one basic vector is our
$\frac{``\lambda_8``}{2}$, then for $A_{\mu}^{``8``}$ we have the
covariant derivation
$\partial_{\mu} - g_N\,\frac{``\lambda_8``}{2}\,A_{\mu}^{``8``}$.
If this covariant derivative is written in terms of the U(1)--subgroup,
corresponding to monopoles with the Dirac relation (\ref{13a}),
then the covariant derivative has a form
$\partial_{\mu} - g_{(1)}\,A_{\mu}^8\cdot{\un{\un{M}}}$. Here
${\un{\un{M}}}$ has the property that $\exp(i2\pi{\un{\un{M}}})$
corresponds to the elements of the group $SU(N)/Z_N$ going all around and
back to the unit element. Of course, ${\un{\un{M}}}
=\frac{g_N}{g_{(1)}}\cdot\frac{``\lambda_8``}{2}$ and the ratio
$g_N/g_{(1)}$ must be such one that
$\exp(i2\pi\frac{g_N}{g_{(1)}}\frac{``\lambda_8``}{2})$
shall represent --- after first return --- the unit element of the
group $SU(N)/Z_N$. Now this unit element really means the coset consisting
of the center elements $\exp(i\frac{2\pi k}{N})\in SU(N),\,(k\in Z)$, and
the requirement of the normalization of $g_{(1)}$ ensuring the Dirac
relation (\ref{13a}) is:
\be
    \exp(i2\pi\frac{g_N}{g_{(1)}}\frac{``\lambda_8``}{2})
                      = \exp(i\frac{2\pi}{N}){\bf 1\!\!\!1}.
                                           \lb{18a}
\ee
This requirement is satisfied if the eigenvalues of
$\frac{g_{N}}{g_{(1)}}\frac{``\lambda_8``}{2}$ are
modulo 1 equal to $-\frac{1}{N}$, i.e. formally we might write:
\be
   \frac{g_{N}}{g_{(1)}}\frac{``\lambda_8``}{2}
     = - \frac{1}{N}\quad (\rm{mod}\, 1).                  \lb{19a}
\ee
According to (\ref{1a}), we have:
\be
\frac{g_N}{g_{(1)}}\cdot \frac {1}{\sqrt{2N(N-1)}}
\left(\begin{array}{*{4}{c}}
N-1 & 0 & \cdots & 0\\
0 & -1 & \cdots & 0 \\
\vdots & \vdots &\ddots & \vdots\\
0 & 0 &  \cdots & -1 \\
\end{array}
\right)
= - \frac{1}{N}\quad (\rm{mod}\, 1),                  \lb{20a}
\ee
what implies:
\be
   \frac{g_N}{g_{(1)}} = \sqrt{\frac{2(N-1)}{N}},         \lb{21a}
\ee
or
\be
   \frac{g_N^2}{g_{(1)}^2} = \frac{2(N-1)}{N}.         \lb{22a}
\ee

\section{The relation between U(1) and SU(N) critical couplings.\\
The comparison with the Multiple Point Model results}

Collecting the relations (\ref{22a}), (\ref{13a}) and (\ref{3a}),
we get:
$$
   \alpha_N^{-1} = \frac{4\pi}{g_N^2}
= \frac{N}{2(N-1)}\cdot \frac
    {4\pi}{g^2_{(1)}}
= \frac{N}{2(N-1)}\cdot \frac{{\tilde g}^2_{genuine}}{\pi}
$$
\be
= \frac{N}{2(N-1)}\sqrt{N^2-1}\cdot \frac{{\tilde g}^2_{eff}}{\pi}
= \frac{N}{2(N-1)}\sqrt{N^2-1}\cdot \frac{4\pi}{g_{U(1)}^2}
= \frac{N}{2}\sqrt{\frac{N+1}{N-1}}\cdot \alpha_{U(1)}^{-1},
                                                            \lb{23a}
\ee
where
\be
          g_{U(1)}{\tilde g}_{eff} = 2\pi                 \lb{24a}
\ee
and $\alpha_{U(1)}=g_{U(1)}^2/{4\pi}$.

The meaning of this relation is that provided that we have
${\tilde g}_{eff}$ the same for $SU(N)/Z_N$ and U(1) gauge theories
the couplings are related according to Eq.(\ref{23a}).

We have a use for this relation when we want to calculate the phase
transition couplings using the scalar monopole field responsible for
the phase transition in the gauge groups $SU(N)/Z_N$. Having in mind
the "Abelian" dominance in the SU(N) monopole vacuum, we must think
that ${\tilde g}_{eff}^{crit}$ coincides with $g_{crit}$ of the
U(1) gauge theory. Of course, here we have an approximation taking
into account only monopoles interaction and ignoring the relatively
small selfinteractions of the Yang--Mills fields. In this approximation
we obtain the same phase transition (triple point, or critical)
${\tilde g}_{eff}$--coupling which is equal to $g_{crit}$ of U(1)
whatever the gauge group SU(N) might be. Thus we conclude that
for the various groups $U(1)$ and $SU(N)/{Z_N}$, according to Eq.(\ref{23a}),
we have the following relation between the phase transition couplings:
\be
      \alpha_{N,crit}^{-1}
           = \frac{N}{2}\sqrt{\frac{N+1}{N-1}}
                          \alpha_{U(1),crit}^{-1}.
                                            \lb{25a}
\ee
Using the relation (\ref{25a}), we obtain:
\be
    \alpha_{Y,crit}^{-1} : \alpha_{2,crit}^{-1} : \alpha_{3,crit}^{-1}
           = 1 : \sqrt{3} : 3/\sqrt{2} = 1 : 1.73 : 2.12.
                                                                  \lb{26a}
\ee
Let us compare now these relations with the Multiple Point Model results
(\ref{53}). For $\alpha_{Y,crit}^{-1}\approx 9.2$ Eq.(\ref{26a}) gives:
\be
 \alpha_{Y,crit}^{-1} : \alpha_{2,crit}^{-1} : \alpha_{3,crit}^{-1}
    = 9.2 : 15.9 : 19.5.                                     \lb{27a}
\ee
Here we see that in the framework of errors the result (\ref{27a})
coincides with the AGUT--MPM prediction (\ref{53}).
It is necessary to take into account an approximate description
of the confinement dynamics in the SU(N) gauge theories which was
used in this paper.

We are inclined to think that $\alpha_{crit}$ has approximately
the same value along the whole (first order) phase transition border
between the "Coulomb" and confinement phases on the corresponding lattice
phase diagram ($\beta^{lat}$, $\gamma^{lat}$) (see Eq.(\ref{3}) and
Refs.\ct{10}).  We think that the discrepancy between $\alpha_{crit}$ given
by Eq.(\ref{53}) for U(1) gauge theory and the corresponding result
(\ref{39a}) obtained in the lattice investigations and in our Higgs monopole
model, although they are of the same order, can be explained by different
scales responsible for these phase transitions. In spite of this fact, it
seems that the relation (\ref{25a}) is quite general, independent of the
scale, although it is crude due to approximations considered in this paper.

\section{Conclusions}

In the present paper we have considered the Dual Abelian Higgs Model
(DAHM) of scalar monopoles reproducing a confinement mechanism in
lattice gauge theories. Using the Coleman--Weinberg idea of the RG
improvement of the effective potential \ct{20}, we have considered the
RG improved effective potential in the DAHM with $\beta$--functions
calculated in the two--loop approximation. The phase transition between
the Coulomb--like and confinement phases has been investigated in the
U(1) gauge theory by the method developed in Ref.\ct{19}. As previously,
critical coupling constants were calculated. Performing
the comparison with results of the one--loop approximation obtained
in Ref.\ct{19}, where we have received $\alpha_{crit}\approx 0.17$ and
${\tilde \alpha}_{crit}\approx 1.48$, in this paper we see that the
two--loop approximation gives the coincidence of the critical values of
electric and magnetic fine structure constants
($\alpha_{crit}\approx 0.208$ and ${\tilde \alpha}_{crit}\approx 1.20$)
with the lattice result (\ref{4}). Also comparing the one--loop and
two--loop contributions to beta--functions, we have demonstrated
the validity of the perturbation theory in solution of the phase
transition problem in the U(1) gauge theory.

In the second part of our paper we have used the approximation of
lattice artifact monopoles, being described by scalar "point" monopoles and
represented by the Higgs scalar field, to reproduce the lattice
phase transition couplings not only for the $U(1)$ case but also for the
gauge groups $SU(N)/Z_N$ (N=2,3 especially). Since even the non-Abelian
fields around the "lattice artifact" monopoles are taken as going in
only one direction in the Lie algebra, the monopoles are really Abelian.
The direction in the Lie algebra of their fields are, however, gauge
independent and we used an averaging argumentation ending up with the
relating the couplings corresponding to the biggest dual couplings and
therefore essentially to the phase transition couplings by Eq.(\ref{23a}).
This relation between the phase transition fine structure constants for
the groups $U(1)$ and $SU(N)/Z_N$ is:
\be
      \alpha_{N,crit}^{-1}
           = \frac{N}{2}\sqrt{\frac{N+1}{N-1}}
                          \alpha_{U(1),crit}^{-1}.
                                            \lb{28a}
\ee
The most significant conclusion for MPM, predicting from the
Multiple Point Principle (MPP) the values of gauge couplings being so
as to arrange just the (multiple critical) point where most phases meet,
is possibly that our calculations suggest the validity of an approximate
universality of the critical couplings, in spite of the fact that we are
concerned with first order phase transitions. We have shown (by approximate
success) that one can crudely calculate the phase transition couplings
without using any specific lattice, rather only approximating the
lattice artifact monopoles as fundamental (pointlike) particles
condensing. The details of the lattice --- hypercubic or random, with
multiplaquette terms or without them, etc. --- do not matter for the
value of the phase transition coupling so much. Such an approximate
universality is, of course, absolutely needed if there is any sense
in relating lattice phase transition couplings to the experimental
couplings found in nature. Otherwise, such a comparison would only
make sense if we could guess the true lattice in the right model,
what sounds too ambitious.

It must be admitted though that we only properly compared the Parisi
improved couplings which are not the same ones as the couplings
corresponding to the scale of the monopole mass. The corrections
used were taken from the $U(1)$ case in which one has both the
Parisi improved and the long scale distance phase transition couplings
at disposal.

If we could drive the comparison of the phase transition couplings
for the different groups to a higher accuracy, we might seek
in MPM to make the comparison of different critical
couplings somewhat similar to the studies in the GUT models.

ACKNOWLEDGMENTS:
We are greatly thankful to Y.Takanishi for useful discussions and
help.

One of the authors (L.V.L.) is indebted to the Niels Bohr Institute
for its hospitality and financial support.

Financial support from grants SCI-0430-C (TSTS) and CHRX-CT-94-0621
is gratefully acknowledged by H.B.Nielsen.

We want to thank N.Mankoc--Bor\^{s}tnik and the Slovenian Ministery
of Science and late Professor Plemejl for the meetings we had in the
House of Plemejl in Bled.

Also H.B.N. thanks the Bergsoe Foundation for the prize for popularization.

\newpage
\vspace*{-50mm}
\begin{figure}
\centerline{\epsfxsize=\textwidth \epsfbox{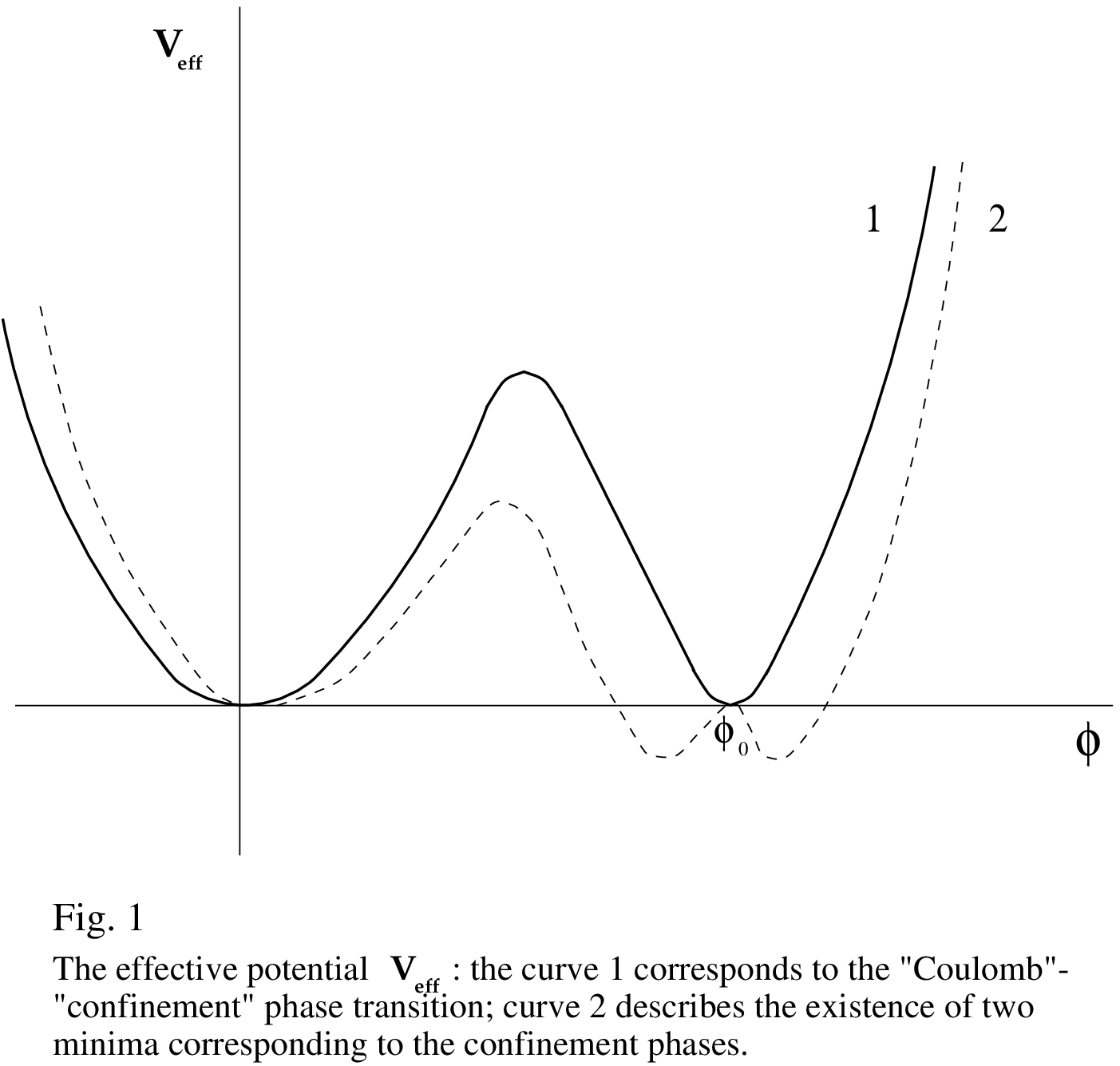}}
\end{figure}

\setcounter{figure}{1}
\newpage
\vspace*{-50mm}
\begin{figure}
\centerline{\epsfxsize=\textwidth \epsfbox{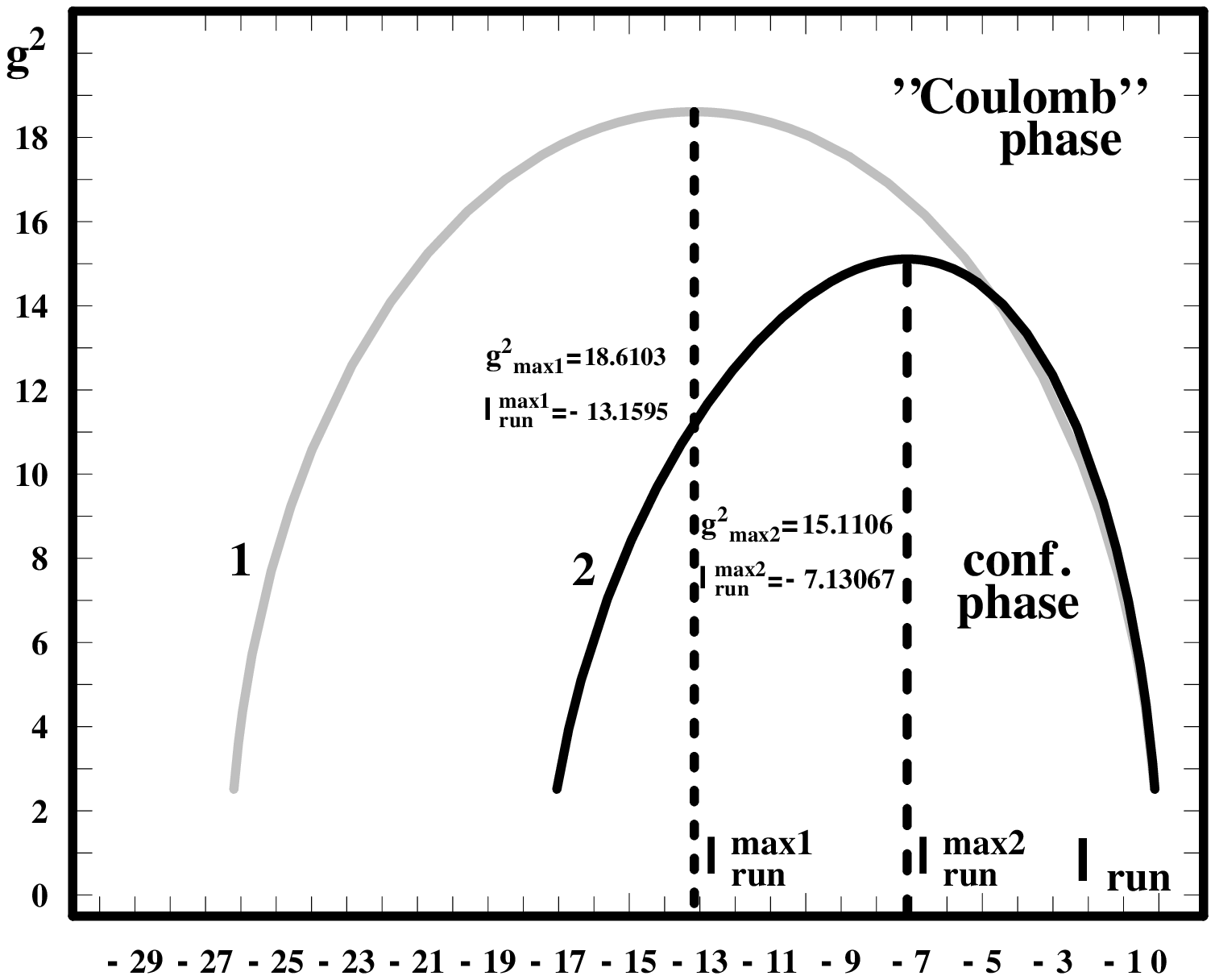}}
\caption{The one-loop (curve 1) and two-loop (curve 2) approximation phase
diagram in the dual Abelian Higgs model.} \end{figure}

\newpage
\vspace*{-50mm}
\begin{figure}
\centerline{\epsfxsize=\textwidth \epsfbox{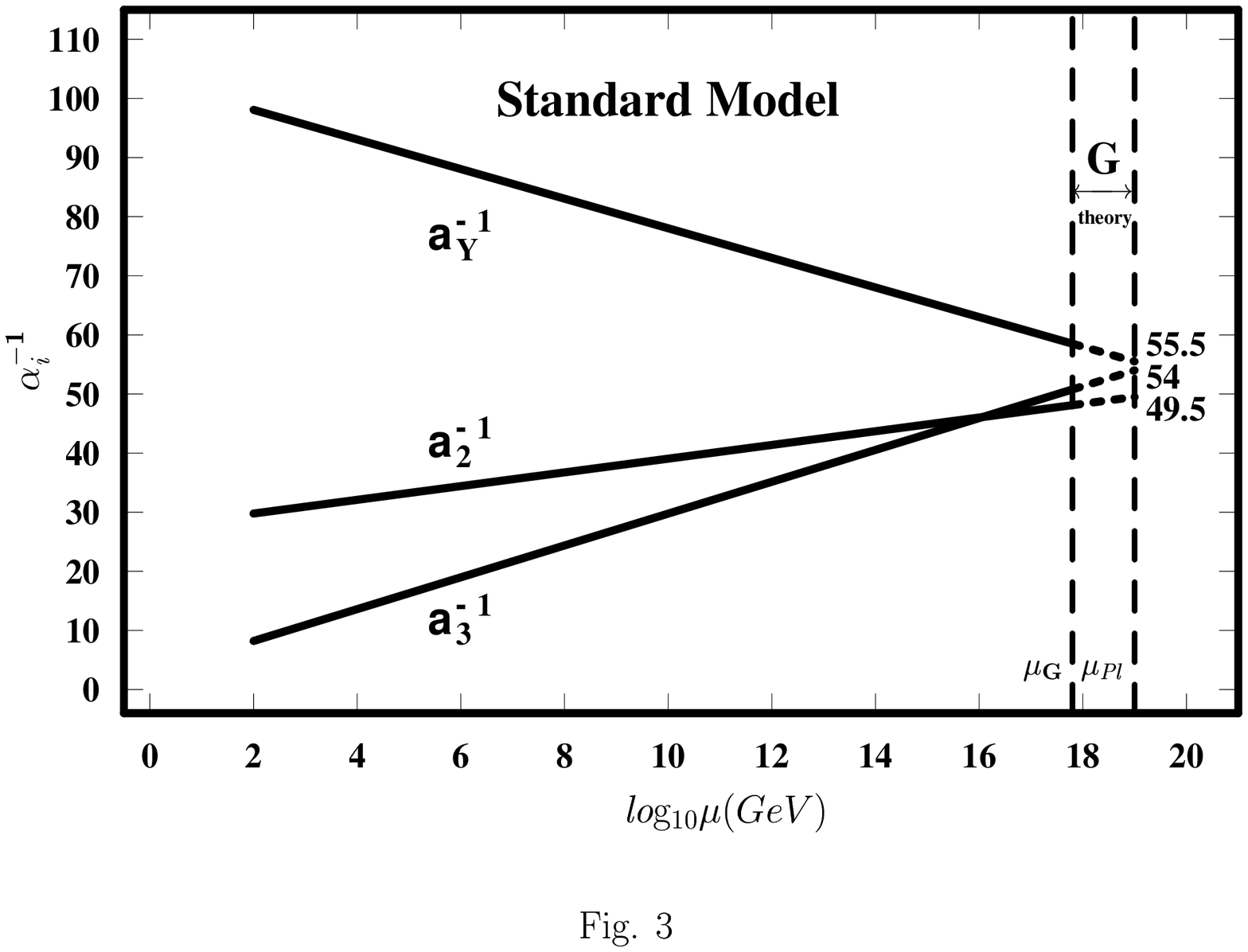}}
\end{figure}

\end{document}